\documentclass[aps,prl,twocolumn,showpacs,amsmath,amssymb,floatfix,superscriptaddress]{revtex4-1}
\usepackage{epsfig} \usepackage{bm}
\usepackage{color}
\usepackage{graphicx}
\usepackage{amssymb}
\usepackage{multirow}

\makeatletter

\@ifundefined{textcolor}{}
{%
 \definecolor{BLACK}{gray}{0}
 \definecolor{WHITE}{gray}{1}
 \definecolor{RED}{rgb}{1,0,0}
 \definecolor{GREEN}{rgb}{0,1,0}
 \definecolor{BLUE}{rgb}{0,0,1}
 \definecolor{CYAN}{cmyk}{1,0,0,0}
 \definecolor{MAGENTA}{cmyk}{0,1,0,0}
 \definecolor{YELLOW}{cmyk}{0,0,1,0}
 }

\makeatother

\usepackage{amsmath}
\usepackage{amsfonts}%
\usepackage{graphicx}
\usepackage{dcolumn}
\usepackage{bm}
\usepackage[mathscr]{eucal}
\usepackage{array}

\newcommand{\req}[1]{Eq.~(\ref{#1})}
\newcommand{\reqs}[1]{Eqs.~(\ref{#1})}
\newcommand{\rref}[1]{(\ref{#1})}

\newcommand{\q}{\mathbf{q}}
\newcommand{\p}{\mathbf{p}}

\renewcommand{\r}{\mathbf{r}}

\newcommand{\beq}{\begin{equation}}
\newcommand{\eeq}{\end{equation}}
\newcommand{\be}{\begin{equation}}
\newcommand{\ee}{\end{equation}}
\newcommand{\beqa}{\begin{eqnarray}}
\newcommand{\eeqa}{\end{eqnarray}}
\newcommand{\bea}{\begin{eqnarray}}
\newcommand{\eea}{\end{eqnarray}}

\usepackage[normalem]{ulem} %
\usepackage{color}

\IfFileExists{srcltx.sty}{\usepackage[active]{srcltx}}

\begin{document}

\title{Tunable Strongly Correlated Band Insulator}

\author{V.V. Cheianov}

\affiliation{Physics Department, Lancaster University, Lancaster, LA1 4YB, UK}

\author{I.L. Aleiner}
\affiliation{Physics Department, Columbia University, New York, NY
  10027, USA }

\affiliation{Physics Department, Lancaster University, Lancaster, LA1 4YB, UK}


\author{V.I. Fal'ko}

\affiliation{Physics Department, Lancaster University, Lancaster, LA1 4YB, UK}

\begin{abstract}
We introduce the notion of the strongly correlated band insulator (SCI), where the lowest energy excitations are collective modes (excitons)
rather than the single particles. We construct controllable $1/N$ expansion for SCI to describe their observables properties. A remarkable example of
the SCI is  bilayer graphene which is shown to be tunable between the SCI and usual weak coupling regime.
\end{abstract}

\maketitle

\begin{figure}[h]
\includegraphics[width=0.84\columnwidth]{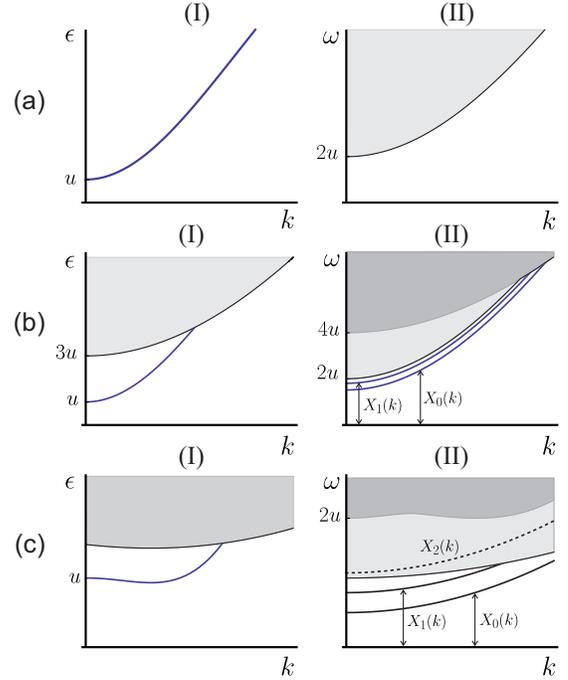}
\caption{
\label{fig1}
One-  (I) and two- particle spectra (II) in the insulator.
a) Non-interacting particles; b) the
weak coupling; c) SCI; Particle-hole symmetry is implied for simplicity.
}
\end{figure}

{\em Introduction} --
Idealized models of band insulators \cite{Kittel} are based upon determining the
spectrum of the single particle excitations (SPEs), Fig.~\ref{fig1}a (I),
which are characterized by the fermionic statistics,
 momentum $k$, charge $\pm e$, spin $1/2$. 
All the other excitations, {\it i.e.}, electron-hole pairs are combined
from the SPEs forming the particle-hole continuum [shaded region on Fig.~\ref{fig1}a (II)].

Weak electron-electron repulsion does not change the gapped spectrum of the SPEs significantly, however  it opens
the decay channel of the SPE into three-particle continuum. This process is allowed only for the particles
with the energy $\epsilon$ above the threshold, see Fig.~\ref{fig1}b (I). 
The qualitative  difference appears in the two-particle spectrum: electron-hole bound
states (excitons) \cite{Wannier} are split down from the particle-hole continuum  Fig.~\ref{fig1}b (II).
Those discrete branches $X_n(k)$ (number of branches is infinite for the interaction potentials 
$\lim_{r\to \infty}U(r)r^2  = \infty$) can not decay unless their energies exceed some threshold.
We will refer to this situation  as {\em weak-coupling insulator}.
All the thermodynamic and transport properties of such insulator are described
by the SPE whereas excitons are responsible for the fine structure of the optical spectra.

With the increase of the interaction the excitation hierarchy in the band insulator changes qualitatively,
see Fig.~\ref{fig1} c.
In this case, $u > X_0(0) $, and all the low temperature thermodynamics and the energy transport
is contributed mostly by the excitons whereas the charge transport is determined by the SPEs.
This leads to the different temperature dependence for the electric and thermal conductivities.
Moreover, unlike in the case of the weak coupling insulators, the SPEs have a very narrow stability range 
(Fig. ~\ref{fig1}c (I)), above which the electron starts producing excitons 
similarly to well-known Schwinger mechanism \cite{Schwinger} of vacuum polarization. For the same reason,  only a finite number of the exciton
branches are stable -- all the other can decay into the two-exciton continuum (Fig.~\ref{fig1}c (II)).
We will call such system  a {\em strongly correlated insulator} (SCI).

In this Letter, we present the case study of the SCI using the bilayer-graphene (BLG)
in a transverse electric field \cite{BLG1,BLG2,BLG3} as an example. A strong motivation
for studying BLG in this context is the possibility to tune it from the 
weak coupling to the SCI, as it is discussed below. 
Using the number of electron species, $N=4$, (two-fold valley and spin degeneracies) as a large parameter we 
obtain analytic results for observables determined by one- and two- particle excitations. 
We emphasize that our study is  general and applicable to any SCI. 

{\em Model}--
The  band structure of  BLG is described by the 
effective single-particle Hamiltonian \cite{Falko} ($\hbar=1$)
\begin{equation}
\hat H_0=\left(\begin{array}{cc}
u_0 & (k_x+i k_y)^2/({2m}) \\
{(k_x-i k_y)^2}/({2m}) & -u_0
\end{array}
\right) 
\label{Hsp}
\end{equation}
where $u_0$ is the interlayer asymmetry tunable by the transverse electric field, 
$m$ is the effective mass, and $\mathbf p = (p_x, p_y)$ is the 
the Bloch momentum counted from the $K$ point of the Brillouin zone. Excitations whose 
momenta are near $K'$ are described by the parity conjugate 
of the Hamiltonian \eqref{Hsp} and form 
another particle species with identical properties.
The four-fold degenerate spectrum of the SPEs corresponding to \req{Hsp} is
\be
\epsilon(k)=\left(u_0^2+{k^4}/({4m^2})\right)^{1/2} \approx u_0 +{k^4}/({8u_0 m^2}).
\label{e}
\ee


We will see that SCI behavior of  BLG is associated with the long range part of the Coulomb interaction $v(\mathbf r)=e^2/r$
so that the effective  Hamiltonian is  
\begin{equation}
H=\int d^2 r \psi^\dagger \hat H_0 \psi +\frac{1}{2} \int d^2 r d^2 r' \rho(\mathbf r) v(\mathbf r-\mathbf r')
\rho(\mathbf r').
\label{H}
\end{equation}
We omitted here the short range interaction with non-trivial matrix structure. Those terms are related to the symmetry breaking in BLG 
and get renormalized with increasing the linear scale \cite{Lemonik}.
 All those renormalizations, as well as the renormalisation of the electron mass and interlayer asymmetry, stop at distances of the order of 
\be
\ell=1/ \sqrt{m u_0},
\label{ell}
\ee
and are incorporated into the SPE spectrum (\ref{Hsp}). 
In \req{H}, $\hat H_0$  given in \req{Hsp} operates on a
two-component vector $\psi(\mathbf r)=[\psi_A(\mathbf r), \psi_B(\mathbf r)]^T$ in the 
sublattice space (summation over the spin and valley indices is implied),
 $\rho = :\psi^\dagger  \psi:$ is the 
normal-ordered particle density.

Dimensional analysis of the Hamiltonian \rref{H} reveals only one dimensionless coupling constant
\begin{equation}
\kappa^2={m e^4}/{ u_0}, 
\label{intpot}
\end{equation}
which diverges 
as $u_0\to 0$ signaling the  SCI. 

\begin{figure}
\includegraphics[width =0.9 \columnwidth]{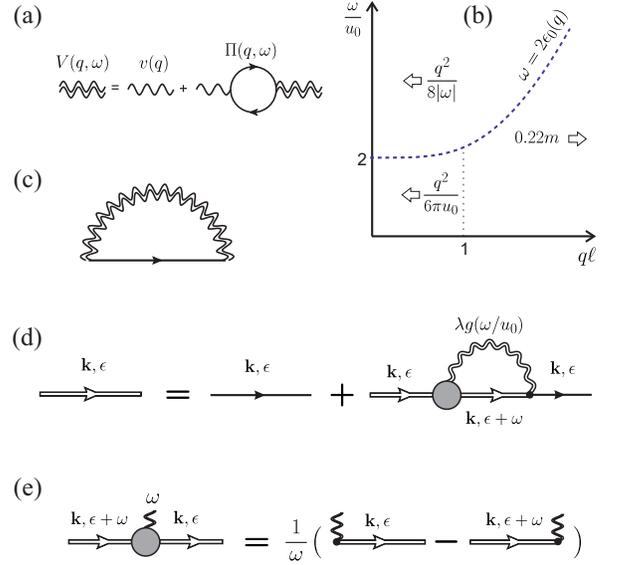}
\caption{a) Resummation of the leading $N$ 
loops leads to the RPA screening of the interaction potential.
b) Asymptotic regions for the polarization operator.
c) The leading-order self energy correction.
d) Resummation of the leading logartihm using Ward-Takahashi identity (e).
\label{polar}
}
\end{figure}

{\em Screened interaction} in large $N$ approximation 
is obtained as a resummation of Fig.~\ref{polar} a:
\be
V(q, i\omega)= {v( q)}
\left[1+N v( q) \Pi(q , 
i\omega)\right]^{-1},
\label{Vqw}
\ee
where $v( q)=2\pi e^2/q$ 
and $\Pi( q, i \omega)$ is the polarization function
in the momentum-frequency domain with 
asymptotic expressions shown in Fig.~\ref{polar} b).
Taking  $N\gg 1$ limit, one finds  
 $\kappa$ to drop out of the 
expression giving
\begin{equation}
V( q, i\omega) \simeq {1}/[{N \Pi( q, i \omega)}].
\label{1overPi}
\end{equation}
Thus,  the perturbation series becomes an expansion 
in powers of $1/N$. Such an expansion, however, contains 
singular terms caused by $\Pi(q \to 0, i\omega) \to 0$.
The expression for the screened static potential is most instructive:
\be
V(r)\equiv \int\!\! \frac{d^2 qe^{i\q\r}}{(2\pi)^2}V(\q,0) =
\left\{
\begin{array}{cc}
 \frac{3 u_0}{N} \ln \frac{R_*}{r} , &  r \ll R_* \\
\frac{e^2}{r}, & r\gg R_* 
\end{array}
\right.,
\label{screenedv}
\ee
where new spatial scale appeared:
\begin{equation}
R_*= {\kappa N\ell}.
\label{IRC}
\end{equation}
Notice that $V(\ell) > u_0$ at at small enough $u_0$, which makes the SCI insulator regime possible.
At larger $u_0$ interaction is weak and may be treated perturbatively.

{\em SPE spectrum and particle residue} --
We start with calculating the leading in $1/N$ self-energy correction, Fig.~\ref{polar}c. 
Finding the  pole of the resulting Green function, we obtain instead of \req{e}
\be
\epsilon(k)=u - \frac{0.680}{N} \frac{k^2}{2 m}
+ \frac{k^4}{8m^2 u_0} \left(1+{\cal O}(1/N)\right).
\label{sps}
\ee
The single particle gap is strongly enhanced:
\be
u=u_0\left(1+\frac{3}{2N} \ln \kappa N\right).
\label{u}
\ee
In the limit of  $\kappa \to \infty$ this enhancement diverges which, as we will see shortly, signals the transition to the SCI.
The physical meaning of gap renormalization is that the band insulator can not completely readjust itself when a charged particle is introduced.
The interaction of the extra particle with the dipoles from the distances between $\ell$ and $R_*$ leads to the logarithmic divergence for $R_*\to\infty$.
The second term in \req{sps} is the interaction-induced negative curvature analytic in $1/N$. 
It significantly exceeds contribution of Ref.~\cite{Falko} to the resulting ``mexican hat'' spectrum
shown on Fig.~\ref{fig1}c (I). 

The single-particle residue  found from the same diagram of  Fig.~\ref{polar}c
is given by
\be
Z\approx 1-({8}/{\pi^2 N}) \ln( \kappa N )  \ln\left( {\Lambda}/{u_0}\right), 
\label{Zapprox}
\ee
where $\Lambda\approx 0.2$ eV is the upper limit for the applicability of the two-band model \rref{Hsp}.

Equations \rref{u} and \rref{Zapprox} contain logarithmically divergent factors, which makes it necessary to  sum up
all orders of the perturbation theory in   
\be
\lambda \equiv \left(\ln \kappa N\right)/(2\pi N).
\label{lambda}
\ee
The summation procedure outlined below shows that the  gap \rref{u} 
does not acquire any higher order in $\lambda$ corrections
whereas \req{Zapprox} changes to
\be
Z=\exp\left[-\left({8}/{\pi^2 N}\right) \ln( \kappa N )  \ln\left( {\Lambda}/{u_0}\right)\right].
\label{Z}
\ee

Let us sketch derivations of \reqs{u} and \rref{Z}. We notice that the dominant logarithmic divergence is contributed by the
momenta transfer $q\ell \lesssim 1$ through any interaction propagator. This  
allows for following simplifications: (i) to factorize the polarization operator as
\be
\Pi(q, i \omega)={q^2}/\left[{u_0 g(i\omega/u_0)}\right],
\label{factorized}
\ee
where $g(z)$ is a function well defined  in the limit $\kappa \to \infty$;
(ii) to neglect the change in the momentum $k$ of any single particle Green function (GF).
The latter allows the integration over the momenta in each interaction line separately:
 the screened interaction \rref{Vqw} is replaced with 
\be
 \int^{q\ell \lesssim 1}  V(q,i\omega){d^2 q}/{(2\pi)^2}
=  u_0 \lambda g\left({i\omega}/{u_0}\right).
\label{replacement}
\ee
A non-trivial aspect of \req{factorized} is the appearance on the bare scale ${u_0}$ rather than the renormalized gap $u$ in
the right-hand-side of the equation.
The reason for this is that divergence $\propto \ln N\kappa$ can not enter into $\Pi$,
because the screening is determined by the neutral dipoles
\cite{footnote}.

Dyson equation for the resulting strong coupling theory, Fig.~\ref{polar}d, is then closed using the Ward-Takahashi identity
Fig.~\ref{polar}e. This results in exact time ordered GF
\bea
&&G(\mathbf k, \epsilon)=\int_{-\infty}^{\infty} d t e^{i \epsilon t} G_0(\mathbf k, t) 
e^{-\frac{\lambda}{2} F(t)}, \label{dre}
\\
&&F(t)=i g(0)  u |t| + \int_2^{\Lambda/u_0} dz \left (1-e^{-i z |t|}\right) W(z).
\nonumber
\eea
where $G_0(\mathbf k,t )$ is the bare GF in the momentum-time domain,
and $W(z)\equiv i \left[g(z+i0)-g(z-i 0)\right]/(\pi  z^2) $.

Expression \rref{dre} is valid for any function $g$.
For the polarization loop neglecting vertex corrections and 
renormalization of GF, $N\to \infty$ \cite{footnote}, 
we find $g(0)=6\pi$ and 
\begin{equation} \notag
W(z)=
\frac{16 \pi  z}{z^2+4} \left[{\pi ^2+\left(\ln \frac{z+2}{z-2}-\frac{4
   z}{z^2+4}\right)^2}\right]^{-1}.
\end{equation}
Then, 
$F(t)= 6\pi i u |t| + (16/\pi)\ln( \Lambda/u_0)$ and \req{dre} leads to \reqs{u} and \rref{Z}.

\begin{figure}
\includegraphics[width =0.8\columnwidth]{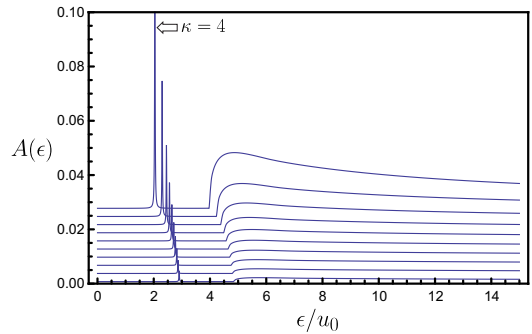}

\caption{The electron spectral weight $A(\epsilon)$ at different values of 
the coupling constant $\kappa$. For clarity, 
the curves corresponding to ten equidistant
values of  $\kappa =4, 8, 12, \dots, 40$ are vertically offset 
and the quasiparticle peak is artificially broadened and scaled down by a factor 
of 100.  
}
\label{A}
\end{figure}

Equations \rref{dre} enables one to calculate not only the particle pole but also the incoherent contribution describing
the coupling of the extra electron introduced or extracted from the system (as in the tunneling or photoemission
experiments) with many particle continuum. To do so, we evaluate 
 $A(\epsilon)=
 (1/\pi) |\mathrm{Im} \mathrm{Tr} G(0, \epsilon)|$ by numerical integration of \req{dre}.
(Due to the electron-hole symmetry, $A(\epsilon)=A(-\epsilon)$, and the structure and finite $k$ is similar).
The result plotted in Fig.~\ref{A} shows the single particle peak at $\epsilon=u\gg u_0$ and the
threshold at $\epsilon= u+2u_0$ due to the coupling to the three-particle continuum \cite{footnote2}

\begin{figure}
\includegraphics[width =0.9 \columnwidth]{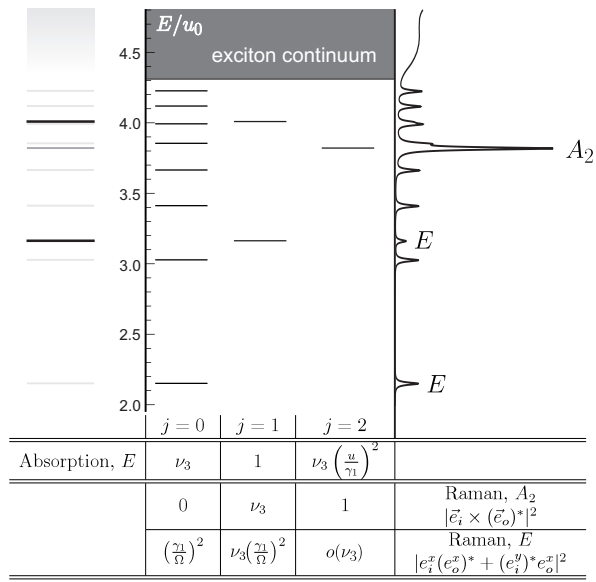}



\caption{The energy levels of the exciton in units of $u_0$ for $N=4$. Hatched region
is the two-exciton continuum.
Side panels are the sketches of the absorption (left) and Raman spectrum (right).
Bottom table are the oscillator streghts of the transitions in units of the strongest line,
$\vec{e}_{i,o}$ are the complex polarization vectors for the incoming (i) and outgoing (o) light,
$\Omega$ is the base line frequency and $\gamma_1 \simeq 0.2$ eV. The coefficient $\nu_3\simeq({\cal E}_{\mathrm{LiTr}}/u)^2$ is not zero only due to
the trigonal warping of the spectrum, where ${\cal E}_{\mathrm{ LiTr}}\simeq 1$ meV is the 
energy of the Lifshitz transition in gapless, $u_0=0$, BLG \cite{Falko,Lemonik}.  
}
\label{exlevels}
\end{figure}

{\em Exciton spectra} -- Next we discuss the collective spectrum of the system,
revealed as the poles of two-particle propagators of the system.
To calculate those poles, $X$, in leading $1/N$ approximation it is sufficient to neglect the retardation
in the interaction potential \rref{Vqw} and consider  Schr\"odinger's equation for an electron and a hole,
\be
\left\{\frac{\hat p_e^4+p_h^4}{8 m^2 u_0}+\left[2  u - V(r)\right]\right\}\Psi(r_e, r_h)= 
X\Psi(r_e, r_h),
\label{SrEq}
\ee
with $V(r)$  defined in \req{screenedv}, and $\r\equiv \r_e-\r_h$.
Unbound states with $X > 2u$ correspond to the particle-hole continuum and the bound states are the exciton lines.

\begin{subequations}
Due to the non-parabolicity of the one-particle spectrum the motion of the exciton center of mass ${\mathbf P}=\p_e+\p_h$ can not be separated from
the relative motion, and we consider here the case of $P=0$ relevant for optics. In this case, the levels are labeled
by the normal angular momenta $|j|=0,1,2, \dots$ and by the radial quantum number $n=0,1,\dots$.
Dimensional analysis gives the low lying states with the size of the order of $\ell \ll R_*$,
thus $\left[2 u - V(r)\right]=u_0\left[2+ (3/2N)\ln r/\ell\right]$ does not depend on $\kappa \gg 1$,
and,  for $n+ |j| \lesssim N\kappa$,
\be
X_n^j=u_0\left[2 + \frac{3}{N}\left(\frac{1}{4}\ln \frac{12}{N} + \xi_n^j\right)\right],
\label{levels}
\ee
where $\xi_n^j$ are the eigenvalues of the differential operators 
\be
\hat{h}^j= \left(-{d^2}/{d z^2}- {z}^{-1}{d}/{d z}+{j^2}/{z^2}\right)^2+ \ln z.
\label{levels2}
\ee
\end{subequations}
Results of the numerical diagonalization of $\hat{h}^j$ are plotted on Fig.~\ref{exlevels}. 
Note that the levels with $X_n^j > 2X^0_0$ decay into two exciton continuum as in Fig.~\ref{fig1}c (II).

Since the spectrum of the excitons \rref{levels} does not depend on the coupling constant $\kappa$, the system can be quite easily
tuned in the SCI regime, $u\gg X_0^0$. 

The fact that energies of the lowest-lying excitons in \req{levels} scale with bare gap parameter $u_0$ 
rather than the renormalised single-particle gap $u$ justifies the scaling form of the polarization operator \req{factorized}.
 Indeed, for the strong coupling regime, the polarisation of BLG occurs 
via appearence of virtual excitons with $j=\pm 1$, so that $\Pi(q\ll\ell^{-1}, \omega)$ is generic for both weak coupling regime and SCI.
Moreover, $g(0)$ acquires only $1/N$ corrections, so that \req{u} does not change. The fine structure of $g(z)$ shows the exciton
resonaces it changes the threshold in $A(\epsilon)$ by the exciton binding energy $\simeq u_0/N$ and introduces an additional fine structure
which may be distinguished in higher derivatives of $A(\epsilon)$

{\em Exciton lines in SCI optics} --
The exciton lines \rref{levels} are degenerate ($N^2$ for j=0, and $2N^2$ for $|j|>0$) due to the $N$-fold degeneracy of the electron spectrum.
Such degeneracy is lifted due to the crystalline symmetry \cite{exchange}.

First, the $4$-fold spin degeneracy is split to the $S=0$ singlet and $S=1$ triplet states due to the exchange  interaction.
In the absence of spin-orbit interaction triplet excitons can not be observed in optical experiments. 
Spin-singlet states are further split due to the trigonal symmetry of the bilayer crystal and should be 
classified according to the irreducible representations of its planar group, only $A_{1,2}$ and $E$ representations are optically active.
Such a classification is presented
in Fig.~\ref{exlevels} together with the selection rules ($A_1$ is not active in Raman because of the electron-hole symmetry). 
These rules for the bright $E$-exciton absorption (as well as luminescence), 
$\sigma^{\pm} \to X^1 \to \sigma^{\pm}$ are determined by the form of the interband 
current operator derived from the Hamiltonian \req{Hsp}, and a trigonal warping term \cite{Falko} 
due to skew interlayer hopping, for a weak transition $\sigma^{\pm} \to X^0$.  
The selection rules for Raman processes are determined by the electron-two photon interaction 
via virtual intermediate state \cite{Kashuba}, with the dominant transition $A_2$: $\sigma^{\pm} \to \pm i \sigma^{\pm}+X^2$ 
and a satelite $E$: $\sigma^{\pm} \to \sigma^{\mp}+X^0$. 

{\em In conclusion,} 
we presented a general and controllable theory of a strongly correlated insulator (SCI): a band insulator where the 
the spectrum of excitons lies deep below the lowest branch of the single-particle spectrum. Gapped bilayer graphene is not the only 
example with such properties, the list of other potential SCIs includes quantum wells of 
semimetal compounds, such as Bi$_{1-x}$Sb$_x$, or silicene 
\cite{silicene} in a transverse electric field \cite{LU-gang}. However, BLG is unique in its
tunability from the weak coupling  to the SCI.  In the lab, this tuning can be achieved by the application of the electric field normal to the BLG plane.

We acknowledge discussions with L.I. Glazman and support by US DOE contract No. DE-
AC02-06CH11357 (IA), EPSRC EP/G041954/1 (VF), ERC (VC), and the Royal Society  (VF).


\begin{thebibliography}{99}
\bibitem{footnote} We will elaborate on this after the exciton is considered.
\bibitem{footnote2}In fact $A(\epsilon)$ is singular at $\epsilon_n=u+ 2n u_0$ but those additional singularities are seen only in higher derivatives.
\end{thebibliography}
\end{document}